\documentclass{edm_template}
\usepackage{multirow}
\usepackage{tabularx}
\newcommand{\elm}{Elo\xspace}
\newcommand{\melm}{M-Elo\xspace}
\newcommand{\name}{RiPPLE\xspace}
\usepackage{xspace}
\renewcommand{\footnotesize}{\scriptsize}

\begin{document}

\title{A Multivariate Elo-based Learner Model\\ for Adaptive Educational Systems }
%
%
%
%
%

\numberofauthors{4} 
%
\author{
%
%
\alignauthor
Solmaz Abdi\\
      \affaddr{The University of Queensland}\\
      \email{solmaz.abdi@uq.edu.au}
\alignauthor
Hassan Khosravi\\
       \affaddr{The University of Queensland}\\
       \email{h.khosravi@uq.edu.au}
\alignauthor 
Shazia Sadiq\\
       \affaddr{The University of Queensland}\\
       \email{shazia@itee.uq.edu.au}
\and  
\alignauthor Dragan Gasevic\\
      \affaddr{Monash University}\\
       \email{dragan.gasevic@monash.edu	}
}


\maketitle
\begin{abstract}
The Elo rating system has been recognised as an effective method for modelling students and items within adaptive educational systems. The existing Elo-based models have the limiting assumption that items are only tagged with a single concept and are mainly studied in the context of adaptive testing systems. In this paper, we introduce a multivariate Elo-based learner model that is suitable for the domains where learning items can be tagged with multiple concepts, and investigate its fit in the context of adaptive learning. To evaluate the model, we first compare the predictive performance of the proposed model against the standard Elo-based model using synthetic and public data sets. Our results from this study indicate that our proposed model has superior predictive performance compared to the standard Elo-based model, but the difference is rather small. We then investigate the fit of the proposed multivariate Elo-based model by integrating it into an adaptive learning system which incorporates the principles of open learner models (OLMs). The results from this study suggest that the availability of additional parameters derived from multivariate Elo-based models have two further advantages: guiding adaptive behaviour for the system and providing additional insight for students and instructors.
\end{abstract}

%

\section{Introduction}
Adaptive educational systems make use of data about students, learning process, and learning products to adapt the level or type of instruction for each student. Commonly, this adaptivity takes the form of selecting items from a large repository of learning resources to match the current learning ability of a student \cite{paramythis2003adaptive}. To do so, adaptive educational systems rely on learner models that capture an abstract representation of a student's ability level based on their performance and interactions with the system \cite{desmarais2012review}.
Two conventional approaches have been heavily studied for modelling learners. (1) Bayesian Knowledge Tracing (BKT) uses a Hidden Markov Model for capturing students' knowledge state as a set of binary variables indicating whether a knowledge component has been mastered \cite{corbett1994knowledge}. (2) Item Response Theory (IRT) \cite{embretson2013item} and its extensions such as Additive Factor Model (AFM) \cite{cen2006learning} and Performance Factor Analysis (PFA) \cite{pavlik2009performance} rely on a logistic regression model to estimate latent traits related to students' knowledge state and the difficulty of learning items. Neither of these approaches, however, can be easily integrated into online adaptive educational systems as they generally require pre-calibration on big samples of data and ongoing addition of new students and new learning items to the system necessitates continuous calibration of model parameters \cite{pelanek2016applications}. 

The Elo rating system has been shown to be an effective alternative to the above mentioned conventional approaches for modelling students in adaptive educational systems \cite{klinkenberg2011computer}. It is simple, fast, robust and order-sensitive which makes it a suitable model for adaptive educational systems where it is required to update students' proficiency level upon administration of each question  \cite{wauters2010monitoring}. In the educational context, the Elo-based model is employed to conduct a paired comparison among students and learning items as competitors \cite{wauters2010monitoring}. This model is self-correcting, meaning that the ratings, in the long run, should correctly reflect students' knowledge states and difficulty levels of questions \cite{pelanek2017Elo}. 

The majority of the existing studies on Elo-based models have the following two characteristics: They (1) use repositories that contain items that are pure \cite{doebler2015adaptive} and are tagged with a single concept and (2) are studied in the context of adaptive testing systems such as computerised adaptive testing (CAT) \cite{de2013theory}. The contribution of this paper lies in (1) introducing a new variant of the Elo-based algorithm called \melm that has the capacity to model students and items using repositories that contain items that are tagged with one or more concepts, and (2) investigating its fit in the context of adaptive learning systems, which in contrast to adaptive testing systems, take a more student-centred approach with the aim of assisting students in their learning. 

To evaluate the applicability of \melm in the context of adaptive learning where learning items are explicitly tagged with one or more concepts, we first compare the predictive performance of \melm against the standard Elo-based model using simulated data sets. 
The results from these experiments demonstrate that the behaviour of the \melm is consistent with expectations over a range of parameter settings, and therefore this provides some evidence to suggest that it will be robust in a real-world setting. We then compare the predictive performance of \melm against the standard Elo-based model using public data sets. The results from the experiments indicate that \melm have superior predictive performance compared to the standard Elo-based model, but the difference is rather small. Finally, we integrate \melm into an adaptive learning system and conduct a case study by piloting this system in a large introductory course on relational databases at The University of Queensland. 

The results obtained from our case study suggests that using \melm, as a multivariate Elo-based model, in adaptive learning systems have two additional advantages beyond the standard Elo-based model: The first advantage lies in the availability of additional parameters that provide insight into the characteristics of the domain and the learning process which in turn can be used for guiding adaptivity. The second advantage is that when \melm is opened based on the concepts of open learner models (OLMs) \cite{bull2010open}, it can provide additional insight on course level and individual level competencies and gaps, that can be used by instructors for improving learning item design, while also providing meta-cognitive benefits for students, such as increased motivation and trust in the system. The conducted case study also revealed some of the shortcomings of using Elo-based models in the context of adaptive learning. The main issue is that Elo-based models consider items and students as identical rivals. This assumption seems to fail to adequately model the long-term use of an adaptive learning system in which students' knowledge may increase over time while the difficulty level of the items remains constant.

\section{Background}\label{sec:background}
\vspace{-10pt}
\paragraph{Elo-Based Learner Models}
The Elo rating system is originally developed to rate chess players and is established based on paired comparison of data where two chess players compete against each other  
 \cite{wauters2010monitoring}. 
In the educational setting, a similar paired comparison can be conducted between a student and a question being attempted by the student. The standard implementation of the Elo-based model in education resembles the Rasch model employed in IRT that models students and questions with a single global parameter \cite{pelanek2017Elo}. 
 An important extension to the Elo-based model in education is the multivariate Elo-based model proposed by \cite{doebler2015adaptive} in the context of psychometrics, where instead of using a global knowledge parameter for students, it uses an overlay model which estimates the competency of a student in each different concept using a separate parameter. A similar model for adaptive practice of facts was later proposed by \cite{pelanek2017Elo}, which had an additional global parameter compared to the multivariate Elo-based model that was used in combination with the concept level parameters in modelling the ability of a student on each concept. Both of these models make the assumption that items are tagged with a single concept, which limits their applicability in domains where items can be tagged with multiple concepts. For example, in a Programming domain, an arbitrary learning item might be associated with both "lists"and "loops" concepts. 
\melm, our proposed model, is an extension over the multivariate Elo-based model that has the capacity to model students and items in the presence of items with multiple concepts.  
\vspace{-15pt}
\paragraph{Adaptive Testing vs. Adaptive Learning}
Adaptivity in educational systems has been investigated broadly both in the context of computerized adaptive testing (CAT) and adaptive learning \cite{paramythis2003adaptive}.
Generally, adaptive testing systems conduct an exam using a sequence of questions that are successively administered with the purpose of maximising the precision of the system's current estimate of the student's ability. The exam is usually terminated once the system has an estimate of the student's ability with a confidence level that exceeds a user-specified threshold.  In contrast, adaptive learning systems such as ALEKS\footnote{https://www.aleks.com/} take a more student-centred approach with the aim of assisting students in their learning. As such, in most adaptive learning systems, students (1) have the opportunity to decide whether or not to engage with the suggested learning items, (2) can spend, theoretically, an infinite amount of time on a learning item or on the system and (3) will receive rich feedback on their learning after engaging with each learning item. In this paper, we examine the fit of our proposed model, and more generally multivariate Elo-based models in the context of adaptive learning.

\vspace{-15pt}
\paragraph{Open Learner Models}
Open learner models (OLMs) are learner models that are externalised and made accessible to students or other stakeholders such as instructors, often through visualisation, as an important means of supporting learning \cite{bull2010open}. They have been integrated into a variety of learning technologies such as learning dashboards \cite{bodily2018open}, intelligent tutoring systems \cite{Ritter2007} and adaptive learning platforms \cite{disalvo2014khan} to help students and instructors in monitoring, reflecting and regulating learning \cite{bull2010open}.  In this paper, we
investigate the benefits of opening Elo-based models, such as \melm in adaptive learning system based on the principles of the OLMs. 
\section{ Elo-based Learner Modelling}\label{sec:model}
In this section, we first define a mathematical notation for describing the models. Section~\ref{sec:ERS} then provides a review of the standard Elo-based model in the educational context. Finally, Section~\ref{sec:1LM-ERS} introduces our proposed variation of the multivariate Elo-based model, which we call \melm. 

In what follows, let  $U_N =\{ u_1 \ldots u_N \}$ denote a set of students who are enrolled in a course on an adaptive educational system, where $u_n$ refers to an arbitrary student. Each course consists of a set of concepts $\Delta_L = \{ \delta_1 \ldots \delta_L \}$, referred to as the domain model, where $\delta_l$ presents an arbitrary concept. In this work, the notion of a concept is based on taxonomies of knowledge components described by \cite{koedinger2012knowledge}. Let $Q_M = \{ q_1 \ldots q_M \}$ present the content model, denoting a repository of learning items that are available to students in a course in the adaptive learning system, where $q_m$ refers to an arbitrary item.  These learning items can be tagged with one or more knowledge components;  $\Omega_{M\times L}$ is a two-dimensional array, where $\omega_{ml}$ is  $1/g$ if item $q_m$ is tagged with $g$ knowledge components including knowledge component $\delta_l$, and 0 otherwise. Let a two-dimensional array $A_{N\times M}$ keep track of students' attempts on the items, where $a_{nm}=1$ indicates that student $u_n$ has answered item $q_m$ correctly, and $a_{nm}=0$ indicates that student $u_n$ has answered item $q_m$ incorrectly.
\subsection{The Standard Elo-based Learner Model (\elm)} \label{sec:ERS}
In the standard Elo-based learner model (\elm), both students and items are considered as identical rivals. \elm assumes one student parameter $\theta_n$ indicating $u_n$'s global proficiency level on the entire domain and one item parameter $d_m$ presenting the difficulty level of item $q_m$. It uses a logistic function to estimate the probability of a correct answer by a student to a given item based on the difference between the student's global proficiency level and the item's difficulty. \cite{pelanek2017Elo}: 
\begin{equation}\label{formula:prob}
P(a_{nm}=1| {\theta}_{n}, d_{m}) = \sigma ({\theta}_{n} - d_{m})
\end{equation}
To account for the guessing effect in the case of multiple choice items with $c$ possible options, Formula \ref{formula:prob} can be easily replaced with a shifted logistic regression using: $ P(a_{nm}=1| {\theta}_{n}, d_{m}) = \frac{1}{c} + (1-\frac{1}{c})*  \sigma ({\theta}_{n} - d_{m})$. once $u_n$ has attempted  $q_m$, the knowledge state of $u_n$ and difficulty level of $q_m$ are simultaneously updated using the following formulas:
\begin{equation}\label{formula:Elo-student}
\theta_{n} := \theta_{n} + K (a_{nm} - P(a_{nm}=1|\theta_{n}, d_{m}))
\end{equation}
\begin{equation}\label{formula:Elo-question}
d_{m} := d_m + K ( P(a_{nm}=1|{\theta}_{n}, d_{m}) - a_{nm})
\end{equation}
, where $K$ is a constant value determining the sensitivity of the estimations based on the student's last attempt. The updates to the student's knowledge state (Formula \ref{formula:Elo-student}) and the difficulty of the item (Formula \ref{formula:Elo-question}) in \elm follows the principles of zero-sum game, in which the sum of gains (loss) to the student's knowledge state and the loss (gain) to the difficulty of the item after the student answers the item turns out to be zero.  
In most extensions of the Elo-based models, in order to get to a stable estimations for the student's knowledge state and item difficulty, $K$ is replaced with an uncertainty function 
\begin{equation}\label{formula:uncertain}
U(n) = \frac{\gamma}{1 + \beta*n}
\end{equation}
, where $\gamma$ and $\beta$ are constant hyper-parameters determining the starting value and slope of changes, respectively, and $n$ indicates the number of prior updates on student's knowledge state or item difficulty \cite{pelanek2017Elo}.

\subsection{Multi-Concept Multivariate Elo-based \\
Learner model (\melm)} \label{sec:1LM-ERS}
In contrast to \elm, where only one parameter is used to model a student's knowledge state on the entire domain,
\melm uses independent parameters to model the student's knowledge state on each individual concept in the domain, and a global parameter for modelling each item. 
As in \elm, for each learning item $q_m$ there is a global difficulty $d_m$ approximating the difficulty level of the item. For students, let a two-dimensional array $\Lambda_{N \times M}$ represents a student's Elo-based learner model, where $\lambda_{nl}$ represents student $u_n$'s knowledge state on concept $\delta_l$, approximating the proficiency level of the student on that certain concept.  To estimate the probability that student $u_n$ answers an item $q_m$ correctly, we first compute $\Bar{\lambda}_{nm} = {\sum_{l=1}^L \lambda_{nl} \times \omega_{ml}}$, which estimates $u_n$'s average competency on concepts that are associated with $q_m$. We then compute the probability of $u_n$ answering $q_m$ correctly using: 
\begin{equation}
P(a_{nm} = 1| \Bar{\lambda}_{nm}, d_{m}) = \sigma (\Bar{\lambda}_{nm} - d_{m})
\end{equation}
 After $u_n$ answers $q_{m}$, the updated estimate of $d_{m}$ is obtained using: 
 \begin{equation}\label{formula:qUpdate}
d_{m} := d_m + K ( P(a_{nm} = 1|\Bar{\lambda}_{nm}, d_{m}) - a_{nm})
\end{equation}
where $K$ can be replaced with an uncertainty function $U(n)$ presented in Formula~\ref{formula:uncertain}. 
To update student $u_{n}$'s Elo ratings based on the given answer to item $q_m$, we update the student's parameter on each concept $\delta_l$ the question is tagged with separately using the following formula:
\begin{equation}\label{formula:studentUpdate}
\lambda_{nl} := \lambda_{nl} + \alpha.K (a_{nm} - P(a_{nm} = 1|\lambda_{nl}, d_{m}))
\end{equation}
where $\alpha$ is a normalisation factor, ensuring that the zero-sum game principles are enforced in the model. As such,  the net change made to the parameters estimating $u_n$'s proficiency level in the concepts that are associated with $q_m$ (computed by Formula~\ref{formula:studentUpdate}) and $d_m$ (computed by Formula~\ref{formula:qUpdate}) sum to zero.
$\alpha$ is computed using the following formula:
\begin{equation}
\alpha = \frac{|P(a_{nm} = 1|\Bar{\lambda}_{nm}, d_{m}) - a_{nm}|}{\sum_{l=1}^L (|a_{nm} - P(a_{nm} = 1|\lambda_{nl}, d_{m}) \times \omega_{ml}|)}
\end{equation}


\section{Evaluation}\label{sec:eval}
In this section, we evaluate the suitability of \melm in the context of adaptive learning systems. Section~\ref{eval:synthetic} compares the predictive performance of \melm against \elm using a suite of simulated data sets. Section~\ref{eval:historicalData} compares the predictive performance of \melm against \elm using publicly available data sets.  As commonly used in the evaluation of learner models, we use the area under the curve (AUC), root mean squared error (RMSE) and accuracy (ACC) for reporting the predictive performance of the models. For all experiments, students' knowledge states and item difficulties in \elm and \melm are initialised to zero. Section~\ref{sec:caseStudy} then reports the results of a case study that integrates \melm into an adaptive learning system which incorporates the principles of OLMs.
%


\subsection{Synthetic Data Sets}\label{eval:synthetic}
Synthetic data sets were used to assess the behaviour of the models under different settings by varying parameters in the data generation template. The synthetic data sets were generated using a sequence of steps proposed by \cite{khosravi2017}. At first, a set of students with predefined knowledge states over a set of knowledge components were created. Assigning students' knowledge state was performed by sampling from a normal distribution, where the mean of distribution for each student was sampled from a uniform distribution. In this model, the standard deviation ($\sigma$) of the normal distribution determined the complexity of a student's knowledge state, where smaller values of $\sigma$ led to having students that had roughly the same ability across all of the knowledge components and bigger values of $\sigma$ led to having students with a higher diversity on their abilities across all of the knowledge components. Then, a set of learning items with pre-defined concepts, level of difficulty and discrimination was generated. Assigning concepts to items was performed by sampling from a discrete uniform distribution, while difficulty and discrimination were sampled from a normal distribution. Lastly, to compute the probability that a student $u_n$ answered a learning item $q_m$ correctly, a 2PL Item Response Theory (IRT) \cite{Drasgow1990} model was used as recommended by \cite{Desmarais2010} using: 
$\frac{1}{1 + e ^ {-{a}_m({\theta}_n - {b}_m)}}$, 
where, ${\theta}_n$ represents a student's average competency on the concepts associated with the item,
and ${b}_m$ and ${a}_m$ were the difficulty level and the discrimination level of the item $q_m$, respectively. 
In all synthetic data sets, 100 students, 1000 learning item and 70,000 answers were sampled completely at random. Each learning item was tagged with one to three concepts. In these experiments, 70\% of data was used for training and the remaining 30\% was reserved for the test. Each experiment was repeated five times and the reported values are the average results across the five runs. 

\begin{figure}[!htb] 
  \includegraphics[width=\linewidth]{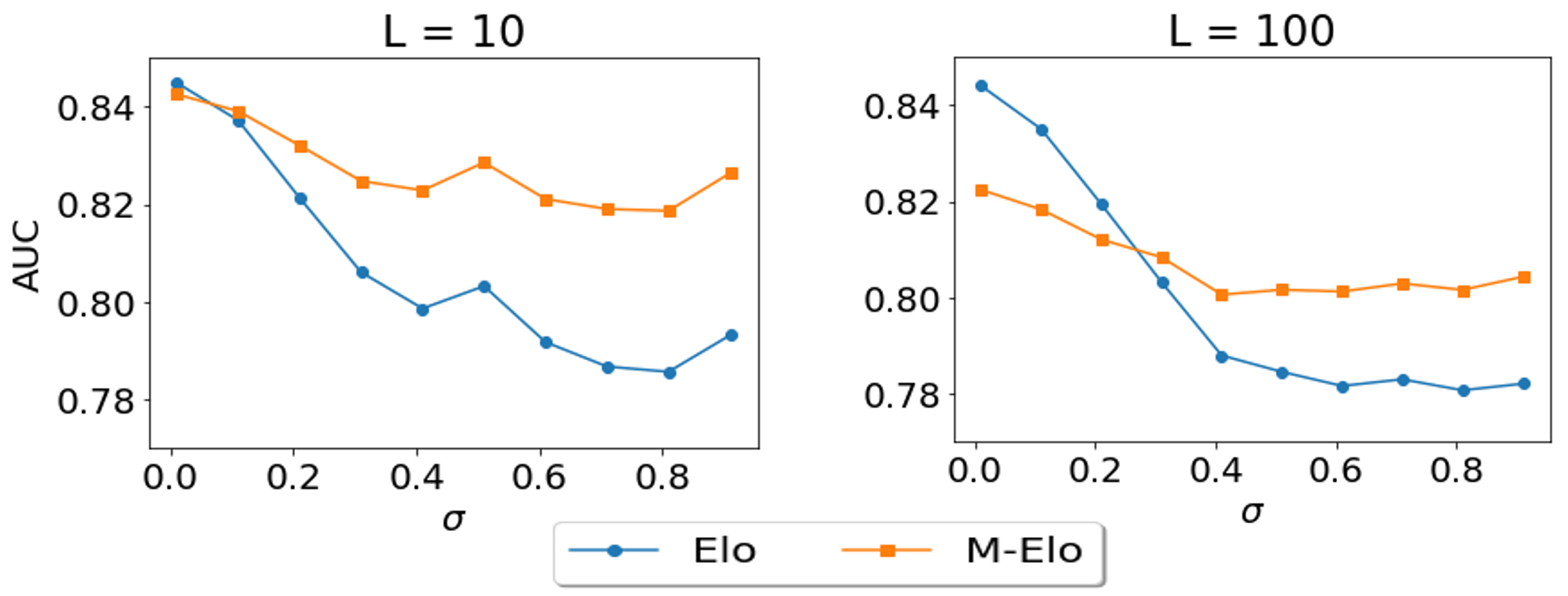}
  \caption{AUC as $\sigma$ is increased. Results are reported for $L=10$,  and $L=100$}\label{fig:alpha}
\end{figure}
Figure \ref{fig:alpha} compares the AUC performance of \elm and \melm in estimating students' knowledge states with respect to different values of $\sigma$ under two different settings for the number of knowledge components ($L$). Our results for $L = 10$ suggest that for smaller values of $\sigma$, \elm outperforms \melm. This was predictable, as in this setting, each student has roughly the same competency level on all of the concepts, and since \elm relies on a global knowledge estimation for students proportional to their overall performance, it can outperform \melm in this setting. As $\sigma$ is increased and students with more diversity in their abilities across different concepts are generated, \melm outperforms \elm as it is able to discriminate between students' abilities on each individual concept, leading to more accurate estimations of students' knowledge state compared to \elm. By increasing $L$ to 100, the same trend is observable; however, the intersection point for $\sigma$ where \melm outperforms \elm becomes bigger, as in this scenario, with the same amount of data, \melm needs to learn many more parameters independently. Evaluations using RMSE and ACC led to very similar patterns as AUC. Therefore, in the interest of space, figures reporting these results have not been included.

\subsection{Public Data Sets}\label{eval:historicalData}
Three data sets namely 'Algebra I 2005-2006' (Alg2005), 'Algebra I 2006-2007' (Alg2006) and 'Bridge to Algebra 2006-2007' (BAlg2006), which were obtained from the PSLC Datashop were used \cite{koedinger2010data}. These data sets were originally obtained from Carnegie Learning's Cognitive Tutor and were made available as the "DevElopment" sets in KDD Cup 2010 \cite{stamper2010data}. Cognitive Tutor provides a fine representation of knowledge components associated with each item. It is a formative assessment tool, where each step taken by a student to answer a problem is considered as an individual interaction. Each learning item (referred to as 'Step' and presented by 'Step Name' in the data sets) is associated with one or more concept (KC) covered in the course. 
We used the train/test split provided by KDD Cup 2010 and discarded interactions with items that have not clearly been tagged with particular concepts. No students were discarded. Overall information about these data sets are presented in Table~\ref{tab:datasets}. A grid search was conducted to determine optimal values for hyperparameters $\gamma$ and $\beta$ of the uncertainty function, described in Formula~\ref{formula:uncertain}. As also reported by \cite{pelanek2017Elo}, the results were not really sensitive to changes from these parameters.  In all the reported experiments on public data sets, $\gamma$ was set to 1.8 and $\beta$ was set to 0.05.
\vspace*{-5mm}
 \begin{table}[h]
\centering
\scriptsize{
\caption{Public data sets \label{tab:datasets}}
\begin{tabular}{|c|c|c|c|c|c|}
\hline
 Data Set & Students & KC &Items &   multi-KC\footnotemark
  &  Interactions\\ 
\hline
Alg2005 & 575& 112  & 147,914 & 51,171 & 609,979  \\
Alg2006 & 1840  & 714 & 319,151 & 21,415  & 1,825,030 \\
BAlg2006 & 1146 &493  & 19,954 & 1,650 & 1,822,697 \\
\hline
\end{tabular}
}
\end{table}
\footnotetext{multi-KC in Table \ref{tab:datasets} indicates the number of items tagged with two or more knowledge components (KCs)}

Table \ref{tab:realdata_results} compares the AUC, RMSE and accuracy (ACC) 
of the model fit statistics related to each model for estimating students' knowledge state. As it is indicated, on all three data sets \melm outperformed \elm in predicting student performance, but the difference was rather small. Considering the insights obtained from the experiments with synthetic data sets, where \melm was outperforming \elm with a small margin, it may be possible to hypothesise that students often have different competency levels on different concepts, but these differences are often not too significant.
\vspace*{-5mm}

\begin{table}[h]
\centering
\scriptsize{
\caption{AUC, RMSE and ACC for public data sets}\label{tab:realdata_results}	
\begin{tabular}{|l|c|c||c|c||c|c|}
\hline
\multicolumn{1}{|c|}{\multirow{2}{*}{Data Set}} & \multicolumn{2}{c||}{AUC} & \multicolumn{2}{c||}{RMSE} &
\multicolumn{2}{c|}{ACC} 
\\ \cline{2-7} 
\multicolumn{1}{|c|}{}                          & Elo         & M-Elo      & Elo         & M-Elo  & Elo         & M-Elo    \\ \hline
Alg2005                                         & 0.726       & 0.750      & 0.392       & 0.385 & 0.787       & 0.79      \\ \hline
Alge2006                                        & 0.687       & 0.695      & 0.394       & 0.390  & 0.784       & 0.797     \\ \hline
BAlg2006                                        & 0.676       & 0.712      & 0.368       & 0.361   &      0.827       & 0.828\\ \hline
\end{tabular}
}
\end{table}
\subsection{Case Study} \label{sec:caseStudy}
\begin {figure*}[h!]
\centering
\includegraphics[width=11.5 cm]{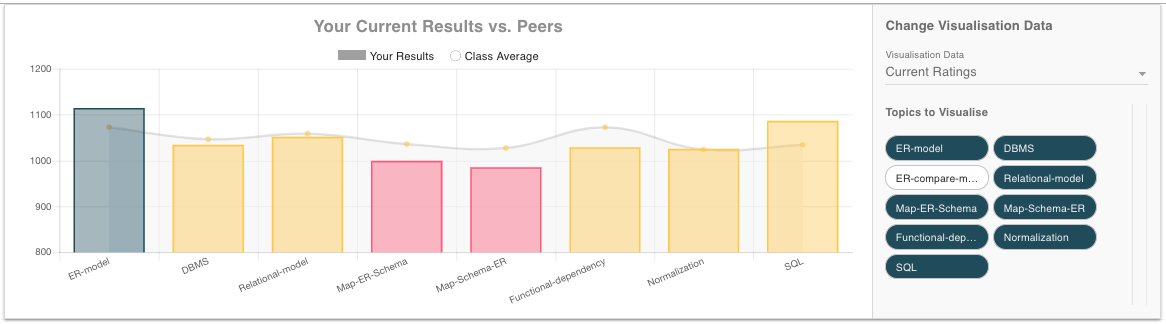}
\caption{Overview of the OLM visualising the current knowledge state of a student.\label{fig:currentrating}} 	
\end {figure*}

To investigate the fit of \melm for adaptive learning in an authentic environment, we integrate \melm into an adaptive learning system called \name \cite{khosraviRiPPLE} and piloted the platform in an introductory course on relational databases at The University of Queensland. The course covers many concepts that are generally included in an introductory course on relational databases including conceptual database design using ER diagrams, relational models, functional dependencies, normalisation, relational algebra, Structured Query Language (SQL), data warehousing and database security. The platform was used for 13 weeks; during this period, 521 of the students enrolled in this course made 91,340 attempts on 1,632 learning items which were available in the system. Among these items, 144 items were tagged with two or more of the 17 concepts, which were associated with the course. Our aim was to investigate the benefits and shortcomings of using \melm in an adaptive system where the model is shared with the students based on the principles of OLMs through a visualisation widget, as indicated in figure \ref{fig:currentrating}, which allows students to visually see their current knowledge state on all concepts of the course. This visualisation is updated and represented to students as soon as they answer a new item from the item pool.

\begin{figure}[!htb] 
  \includegraphics[width=\linewidth ]{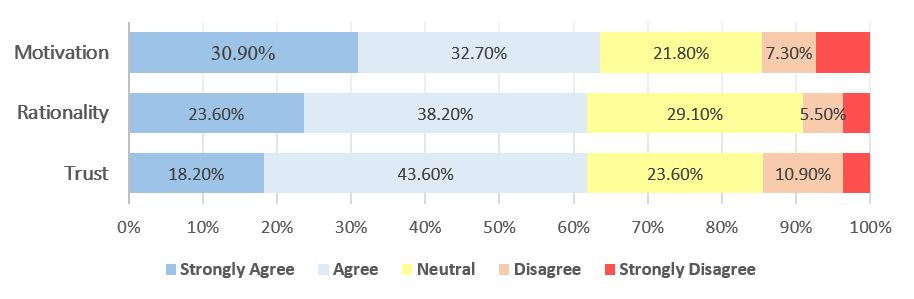}
  \caption{Student survey results}\label{fig:survey}
\end{figure}

\vspace*{-5mm}
\paragraph{Guiding adaptivity}To guide the adaptivity of the system, the estimated knowledge states of the students and difficulties of the questions, as computed by \melm, are passed to the recommendation engine of the adaptive system. For each student, the recommendation engine recommends questions that are on concepts where a student has the largest knowledge gap at a difficulty level that reflect the current learning ability of the student. This adaptivity at a concept-level is because of the availability of the additional parameters learned by \melm, which is not possible to achieve using the standard Elo rating. 
\vspace{-15pt}
\paragraph{Insights from student feedback} To capture students' perspectives about the model, a survey was conducted at the end of the semester using the following three statements: (1) Motivation: the visualisation used by \name for showing my knowledge state increases my motivation to study or further use the system, (2) Rationality: having the ability to visually see my current knowledge state, helps me to understand the rationale behind suggestions made by the system, (3) Trust: having the ability to visually see my  knowledge sate, increases my trust in recommended questions. Responses  were captured using the Likert scale, where 1 represents ``strongly disagree" and 5 represents ``strong agree". The survey also had an open-ended question asking students for feedback on the system. Overall, 55 students who had enrolled in the course and used the system voluntarily participated in the survey. Figure \ref{fig:survey} represents the results of the survey.
The majority of students (63.6\%) agreed or strongly agreed that visualisation of their learner model in \name increased their motivation to further use the system. Furthermore, the majority of the students (61.8\%) agreed that having the ability to visually see their current knowledge state helped them to understand the rationale behind recommendations made by the system and that it increased their trust in recommendations. In their written comments one student mentioned that ``I have used a similar platform in the past, however, the visualisation of my knowledge state in this platform is a great improvement on those.'' Interestingly, two students mentioned that they lost motivation in answering easy questions as the potentially large loss of rating in answering the question incorrectly outweighed the small rating gain received in answering the question correctly. Upon closer examination, we noticed that it seemed challenging to maintain a  balance between the students' proficiency level and the learning items'  difficulty level in our pilot.   Throughout the semester, the average difficulty level of the learning items was falling and the average rating of the students' proficiency was rising. This can be explained by the student-centred design of the system that provided students full access to the internet, textbooks and colleagues as well as an infinite amount of time for answering a question. This may suggest that the zero-sum game principles where the net change in ratings after a student has answered a question is zero might not be ideal for adaptive learning systems. We believe that this may not be an issue for adaptive testing, where the exam-like setting of the system might balance between the students' ratings and the learning items' ratings. 
\vspace{-15pt}
 \paragraph{Insights from instructor feedback} To capture the teaching staff's perspectives, we held informal discussion sessions with members of the teaching staff. They appreciated the additional insight on course level and individual level competencies and gaps that was provided by \melm; similar benefits have also been reported by \cite{pelanek2016applications}. For example, the learning model presented in Figure~\ref{fig:currentrating} indicates that at a class level, students have performed best on ``ER-models" and ``SQL" and have performed worst on ``Map-ER-Schema" and ``Map-Schema-ER" in the course presented in this pilot. A shortcoming that the teaching team has noticed was that students tended to get discouraged from using the system once they had used it for a while as they were not able to make significant changes to their knowledge state despite answering questions correctly. Upon closer examination, we realised that this is due to the use of the uncertainty function, as described in section~\ref{sec:model} and Formula~\ref{formula:uncertain}, which reduces the sensitivity of the estimations as the number of attempts is increased. This functionality seems to be better suited for adaptive testing rather than adaptive learning systems because of the following reason: Given that a student would often take an adaptive test in one sitting with a short timeline and receive no feedback on their work, it is common for adaptive testing systems to assume no learning has occurred during the exam, and as such use of an uncertainty function can help with stabilising the ratings; however,  a student would often interact with an adaptive learning environment over a period of time and competencies might improve via receiving rich feedback on their learning or decline as a result of forgetting. As a consequence, adaptive systems commonly expect the knowledge state of the student to change significantly as they interact with the system. This means that reducing the sensitivity of the estimations over time may restrict the model from unwaveringly evolving to accurately represent the current knowledge state.

\section{Conclusions and Future Directions}\label{sec:discussion}
The aim of this paper was to introduce a new multivariate Elo-based learner model called multi-concept multivariate Elo rating system (\melm) where learning items can be tagged with one or more concepts and investigate its benefits and shortcomings in the context of adaptive learning.
The results from experiments using a suite of synthetic data sets demonstrate that the behaviour of \melm is consistent with expectations and outperforms the standard Elo-based model (\elm) in parameter settings that better reflect real-world environments. The results from experiments on multiple benchmarking public data sets indicate that \melm has slightly superior predictive performance compared to \elm. Conducting a case study suggested that using \melm in adaptive learning systems has additional advantages beyond using \elm.  The first advantage lies in the ability of the model in estimating concept-level competencies which can be used for guiding adaptivity. The second advantage lies in the ability of \melm to be opened based on the principles of OLMs. Making \melm accessible to students increases their motivation to use the platform and increases their trust in the recommendations provided by the platform. It also provides additional insight for instructors on individual student-level or class-level gaps and competencies that can be used to improve item and course design.
The conducted case study gave additional insights into the adverse effects of the zero-sum game design of the Elo-based models as well as using an uncertainty functions in the context of adaptive learning. Future work aims to investigate possible extensions of Elo-based models using these considerations and evaluate their fit for adaptive learning systems. In addition, quantitative comparison of \melm and traditional models of learner models such as AFM and BKT is required to determine its competitiveness with traditional models in predicting students' performance.

\bibliographystyle{acm}
\bibliography{Master-references.bib} 

\begin{thebibliography}{10}

\bibitem{bodily2018open}
{\sc Bodily, R., Kay, J., Aleven, V., Jivet, I., Davis, D., Xhakaj, F., and
  Verbert, K.}
\newblock Open learner models and learning analytics dashboards: a systematic
  review.
\newblock In {\em Proceedings of the 8th International Conference on Learning
  Analytics and Knowledge\/} (2018), ACM, pp.~41--50.

\bibitem{bull2010open}
{\sc Bull, S., and Kay, J.}
\newblock Open learner models.
\newblock In {\em Advances in intelligent tutoring systems}. Springer, 2010,
  pp.~301--322.

\bibitem{cen2006learning}
{\sc Cen, H., Koedinger, K., and Junker, B.}
\newblock Learning factors analysis--a general method for cognitive model
  evaluation and improvement.
\newblock In {\em International Conference on Intelligent Tutoring Systems\/}
  (2006), Springer, pp.~164--175.

\bibitem{corbett1994knowledge}
{\sc Corbett, A.~T., and Anderson, J.~R.}
\newblock Knowledge tracing: Modeling the acquisition of procedural knowledge.
\newblock {\em User modeling and user-adapted interaction 4}, 4 (1994),
  253--278.

\bibitem{de2013theory}
{\sc De~Ayala, R.~J.}
\newblock {\em The theory and practice of item response theory}.
\newblock Guilford Publications, 2013.

\bibitem{desmarais2012review}
{\sc Desmarais, M.~C., and Baker, R.~S.}
\newblock A review of recent advances in learner and skill modeling in
  intelligent learning environments.
\newblock {\em User Modeling and User-Adapted Interaction 22}, 1-2 (2012),
  9--38.

\bibitem{Desmarais2010}
{\sc Desmarais, M.~C., and Pelczer, I.}
\newblock On the faithfulness of simulated student performance data.
\newblock In {\em Educational Data Mining 2010\/} (2010).

\bibitem{disalvo2014khan}
{\sc DiSalvo, B. B. M.~B., et~al.}
\newblock Khan academy gamifies computer science.
\newblock In {\em Proc. 45th ACM Techn. Symp. Comput. Sci. Educ.\/} (2014),
  pp.~39--44.

\bibitem{doebler2015adaptive}
{\sc Doebler, P., Alavash, M., and Giessing, C.}
\newblock Adaptive experiments with a multivariate elo-type algorithm.
\newblock {\em Behavior research methods 47}, 2 (2015), 384--394.

\bibitem{Drasgow1990}
{\sc Drasgow, F., and Hulin, C.~L.}
\newblock Item response theory.

\bibitem{embretson2013item}
{\sc Embretson, S.~E., and Reise, S.~P.}
\newblock {\em Item response theory}.
\newblock Psychology Press, 2013.

\bibitem{khosravi2017}
{\sc Khosravi, H., Cooper, K., and Kitto, K.}
\newblock Riple: Recommendation in peer-learning environments based on
  knowledge gaps and interests.
\newblock {\em JEDM-Journal of Educational Data Mining 9}, 1 (2017), 42--67.

\bibitem{khosraviRiPPLE}
{\sc Khosravi, H., Kitto, K., and Williams, J.~J.}
\newblock Ripple: A crowdsourced adaptive platform for recommendation of
  learning activities.
\newblock {\em preprint arXiv 1910.05522\/} (2019).

\bibitem{klinkenberg2011computer}
{\sc Klinkenberg, S., Straatemeier, M., and van~der Maas, H.~L.}
\newblock Computer adaptive practice of maths ability using a new item response
  model for on the fly ability and difficulty estimation.
\newblock {\em Computers \& Education 57}, 2 (2011), 1813--1824.

\bibitem{koedinger2010data}
{\sc Koedinger, K.~R., Baker, R.~S., Cunningham, K., Skogsholm, A., Leber, B.,
  and Stamper, J.}
\newblock A data repository for the edm community: The pslc datashop.
\newblock {\em Handbook of educational data mining 43\/} (2010), 43--56.

\bibitem{koedinger2012knowledge}
{\sc Koedinger, K.~R., Corbett, A.~T., and Perfetti, C.}
\newblock The knowledge-learning-instruction framework: Bridging the
  science-practice chasm to enhance robust student learning.
\newblock {\em Cognitive science 36}, 5 (2012), 757--798.

\bibitem{paramythis2003adaptive}
{\sc Paramythis, A., and Loidl-Reisinger, S.}
\newblock Adaptive learning environments and e-learning standards.
\newblock In {\em Second european conference on e-learning\/} (2003), vol.~1,
  pp.~369--379.

\bibitem{pavlik2009performance}
{\sc Pavlik~Jr, P.~I., Cen, H., and Koedinger, K.~R.}
\newblock Performance factors analysis--a new alternative to knowledge tracing.
\newblock {\em Online Submission\/} (2009).

\bibitem{pelanek2016applications}
{\sc Pel{\'a}nek, R.}
\newblock Applications of the elo rating system in adaptive educational
  systems.
\newblock {\em Computers \& Education 98\/} (2016), 169--179.

\bibitem{pelanek2017Elo}
{\sc Pel{\'a}nek, R., Papou{\v{s}}ek, J., {\v{R}}ih{\'a}k, J., Stanislav, V.,
  and Ni{\v{z}}nan, J.}
\newblock Elo-based learner modeling for the adaptive practice of facts.
\newblock {\em User Modeling and User-Adapted Interaction 27}, 1 (2017),
  89--118.

\bibitem{Ritter2007}
{\sc Ritter, S., Anderson, J.~R., Koedinger, K.~R., and Corbett, A.}
\newblock Cognitive tutor: Applied research in mathematics education.
\newblock {\em Psychonomic bulletin \& review 14}, 2 (2007), 249--255.

\bibitem{stamper2010data}
{\sc Stamper, J., Niculescu-Mizilm, A., Ritter, S., Gordon, G., and Koedinger,
  K.}
\newblock Data set from kdd cup 2010 educational data mining challenge, 2010.

\bibitem{wauters2010monitoring}
{\sc Wauters, K., Desmet, P., and Van~Noortgate, W.}
\newblock Monitoring learners' proficiency: weight adaptation in the elo rating
  system.
\newblock In {\em Educational Data Mining 2011\/} (2010).

\end{thebibliography}
\end{document}